Proceedings of the 3rd Workshop on Domain-Specific Language Design and Implementation

# DSLDI'15

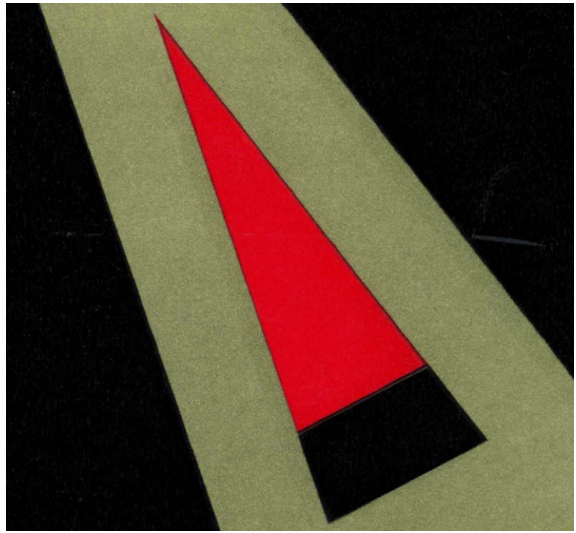

Tijs van der Storm
Centrum Wiskunde & Informatica (CWI)
storm@cwi.nl

Sebastian Erdweg
TU Darmstadt
erdweg@informatik.tu-darmstadt.deJuly 7th, 2015

# DSLDI'15

## Introduction

DSLDI'15 is the 3rd workshop on *Domain-Specific Language Design and Implementation*, which was held at ECOOP 2015, on Tuesday, July 7th, 2015.

DSLDI'15 was organized by:

- Tijs van der Storm
  Centrum Wiskunde & Informatica (CWI)
  `storm@cwi.nl`

- Sebastian Erdweg
  TU Darmstadt
  `erdweg@informatik.tu-darmstadt.de`

The submitted talk proposal were reviewed by the following program committee:

- Emilie Balland
- Martin Bravenboer (LogicBlox)
- Hassan Chafi (Oracle Labs)
- William Cook (UT Austin)
- Shriram Krishnamurthi (Brown University)
- Heather Miller (EPFL)
- Bruno Oliveira (University of Hong Kong)
- Cyrus Omar (CMU)
- Richard Paige (University of York)
- Tony Sloane (Macquarie University)
- Emma Söderberg (Google)
- Emma Tosch (University of Massachusetts, Amherst)
- Jurgen Vinju (CWI)

The website of DSLDI'15 is: `http://2015.ecoop.org/track/dsldi-2015-papers`.

## Informal Post-Proceedings DSLDI'15

This document contains informal post-proceedings of DSLDI'15. It contains:

- A snapshot of the home page of DSLDI'15
- The detailed program of the workshop
- The accepted talk proposals.
- A summary of the panel discussion on *Language Composition*.

# DSLDI 2015

About | Program

## Workshop Goal

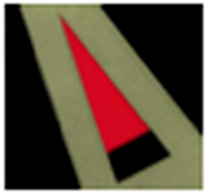

The goal of the DSLDI workshop is to bring together researchers and practitioners interested in sharing ideas on how DSLs should be designed, implemented, supported by tools, and applied in realistic application contexts. We are both interested in discovering how already known domains such as graph processing or machine learning can be best supported by DSLs, but also in exploring new domains that could be targeted by DSLs. More generally, we are interested in building a community that can drive forward the development of modern DSLs.

## Workshop Format

DSLDI is a single-day workshop and will consist of a series of short talks whose main goal is to trigger exchange of opinion and discussions. The talks should be on the topics within DSLDI's area of interest, which include but are not limited to the following ones:

- DSL implementation techniques, including compiler-level and runtime-level solutions
- utilization of domain knowledge for driving optimizations of DSL implementations
- utilizing DSLs for managing parallelism and hardware heterogeneity
- DSL performance and scalability studies
- DSL tools, such as DSL editors and editor plugins, debuggers, refactoring tools, etc.
- applications of DSLs to existing as well as emerging domains, for example graph processing, image processing, machine learning, analytics, robotics, etc.
- practitioners reports, for example descriptions of DSL deployment in a real-life production setting

### DSLDI Summerschool

Are you a student interested in DSL design and implementation? Please consider to also attend the DSLDI summerschool in Lausanne, right after ECOOP! More information here:

http://vjovanov.github.io/dsldi-summer-school/

# Program

**10:05 - 10:20: DSLDI - Welcome** at Karlstejn
Chair(s): Tijs van der Storm, Sebastian Erdweg

| 10:05 - 10:20 Day opening | **Introduction** Sebastian Erdweg, Tijs van der Storm |

**10:20 - 11:20: DSLDI - Session 1** at Karlstejn

| 10:20 - 10:50 Talk | **SCROLL - A Scala-based library for Roles at Runtime** Max Leuthäuser |
| 10:50 - 11:20 Talk | **A case for Rebel, a DSL for product specifications** Jouke Stoel |

**11:30 - 12:30: DSLDI - Session 2** at Karlstejn

| 11:30 - 12:00 Talk | **Flick: A DSL for middleboxes** Nik Sultana |
| 12:00 - 12:30 Talk | **Towards a Next-Generation Parallel Particle-Mesh Language** Sven Karol, Pietro Incardona, Yaser Afshar, Ivo Sbalzarini, Jeronimo Castrillon |

**13:30 - 14:30: DSLDI - Session 3** at Karlstejn

| 13:30 - 14:00 Talk | **DSLs for Graph Algorithms and Graph Pattern Matching** Oskar van Rest, Sungpack Hong, Hassan Chafi |
| 14:00 - 14:30 Talk | **DSLs of Mathematics, Theorems and Translations** Cezar Ionescu, Patrik Jansson |

**14:40 - 15:40: DSLDI - Session 4** at Karlstejn

| 14:40 - 15:10 Talk | **Check Syntax: An Out-of-the-Box Tool for Macro-Based DSLs** Spencer Florence, Ryan Culpepper, Matthew Flatt, Robby Findler |
| 15:10 - 15:40 Talk | **Dynamic Compilation of DSLs** Vojin Jovanovic, Martin Odersky |

**16:10 - 16:40: DSLDI - Session 5** at Karlstejn

| 16:10 - 16:40 Talk | **A practical theory of language-integrated query —and— Everything old is new again** Philip Wadler |

**16:40 - 17:40: DSLDI - Discussion** at Karlstejn

| 16:40 - 17:40 Other | **Panel Discussion: Language Composition** Jonathan Aldrich, Matthew Flatt, Laurence Tratt, Andrzej Wasowski, Sebastian Erdweg |

# SCROLL - A Scala-based library for Roles at Runtime

Max Leuthäuser

Technische Universität Dresden
Software Technology Group
`max.leuthaeuser@tu-dresden.de`

*Today's software systems always need to anticipate changing context. New business rules and functions should be implemented and adapted. The concept of role modeling and programming is frequently discussed for decades across many scientific areas. It allows the modeling and implementation of context dependent information w.r.t. dynamically changing context. Hence future software infrastructures have the intrinsic need to introduce such a role concept. Until now the implementation with existing object oriented languages always requires the generation of a specific runtime environment and management code. The expressiveness of these languages is not able to cope with essential role-specific features, such as true delegation or binding roles dynamically. In this work we present how a relatively simple implementation with Scala based on its Dynamic trait allows to augment an object's type at runtime implementing dynamic (compound-) role types. It enables role-based implementations that lead to more reuse and better separation of concerns.*

Currently, only a handful, mostly unusable (e.g. because they are not providing a running compiler or have been abandoned by the developer) role-based programming languages exists. The field of research is highly fragmented, due to the fact that every research area relies on a different set of role-related features [9]. Therefore it is necessary to establish a basic role concept at runtime and build an appropriate tooling around it to make it more useful for developers. A prototypic Scala implementation for roles (SCROLL - SCala ROLes Language[1]) was developed as library approach, enabling the user to specify roles and context dependent behavior. A basic example of smart (autonomous) cars driving around demonstrates part of its capabilities (see page 2). Two persons (class `Person`) and two cars (`Car`) are bound to the roles `Driver`, `Passenger` and `NormalCar`, `SmartCar` respectively. Each role modifies the default behavior implemented in the players classes, as demonstrated at runtime in the console output listing. Dynamic role selection (like selecting an object playing the `Source` location role) supports filtering for attributes and evaluating function call results (e.g. line 17 and 26, right listing).

Internally the following two technical aspects are the most considerable ones. First, making use of the **Dynamic Trait**. All calls to role functions (i.e. functions that are not natively available on the player object) are translated by the compiler using certain rules[2]. These are adjustable resulting in customizable, *dynamic role dispatch*. Second, applying **implicit conversions**. Scala's implicit classes allow for packing in player and role objects to compound dynamic types. All important role features are exposed this way, e.g. adding, removing and transferring roles or accessing role functions and attributes with the +-Operator (e.g. in line 14, right listing).

The following limitation (apart from some role features not implemented yet, see table 1) is the major subject of the future work on SCROLL. The underlying technique (compiler rewrite rules) hides important typing information to the tooling typically used by most developers, i.e. IDEs with debugger and link tracers. Writing plugins for those (e.g. Eclipse, Intellij) overcoming this issue would be one solution and is currently under development.

Finally, it is necessary to investigate how well this implementation blends into coeval approaches. We use a classification scheme established in two successive surveys on role-based modeling and programming languages, namely [9,11]. This revolves 26 classifying features of roles incorporating both the relational and the context-dependent nature of roles (see table 1). The most sophisticated, competing approach so far is ObjectTeams/Java [6]. It allows to override methods of their player by aspect weaving. Besides that, it introduces *Teams* to represent compartments whose inner classes automatically become roles. Supporting both the inheritance of roles and teams leads to family polymorphism [7]. On the downside, it does not support multiple unrelated player types for a role type. In sum SCROLL provides a simple and clean testing playground *in an unmodified Scala* for using roles at runtime.

| Feature [9] | Chameleon 2003 [4] | OT/J 2005 [6] | Rava 2006 [5] | powerJava 2006 [1] | Rumer 2007 [2] | ScalaRoles 2008 [10] | NextEJ 2009 [8] | JavaStage 2012 [3] | SCROLL 2015 |
|---|---|---|---|---|---|---|---|---|---|
| 1. | ■ | ■ | ■ | ■ | ■ | ■ | ■ | ■ | ■ |
| 2. | □ | ⊞ | □ | ⊞ | ■ | □ | ⊞ | □ | □ |
| 3. | ■ | ■ | ■ | ■ | ■ | ■ | ■ | ■ | ■ |
| 4. | ■ | □ | ■ | ■ | ■ | ■ | ■ | ■ | ■ |
| 5. | ■ | ■ | ■ | ⊞ | ■ | ■ | ■ | ■ | ■ |
| 6. | □ | ■ | □ | ■ | ■ | □ | ■ | □ | □ |
| 7. | ■ | ■ | ■ | ■ | ⊞ | ■ | ■ | ■ | ■ |
| 8. | □ | ■ | □ | ■ | ■ | □ | ■ | ■ | ■ |
| 9. | ■ | ■ | ■ | ■ | ■ | ■ | ■ | ■ | ■ |
| 10. | ■ | ■ | ■ | ■ | ■ | ■ | ■ | ■ | ■ |
| 11. | ■ | ■ | ■ | ■ | ■ | ■ | ■ | ■ | ■ |
| 12. | ■ | ■ | ■ | ■ | ■ | ■ | ■ | □ | ⊞ |
| 13. | □ | ■ | □ | ■ | ■ | ■ | ■ | □ | ■ |
| 14. | ⊞ | ⊞ | □ | ■ | ■ | ■ | ⊞ | □ | ■ |
| 15. | ■ | ■ | ■ | ■ | ■ | ■ | ■ | ■ | ■ |
| 16. | □ | □ | □ | □ | □ | □ | □ | □ | □ |
| 17. | □ | □ | □ | □ | □ | □ | □ | □ | □ |
| 18. | □ | ■ | □ | ⊞ | ⊞ | ⊞ | ■ | □ | ⊞ |
| 19. | □ | ■ | □ | ⊞ | ⊞ | ■ | ■ | □ | ■ |
| 20. | □ | ■ | □ | ■ | ■ | ■ | ■ | □ | ■ |
| 21. | □ | □ | □ | □ | ■ | ⊞ | ■ | □ | ⊞ |
| 22. | □ | ■ | □ | ■ | ■ | ■ | ■ | □ | ■ |
| 23. | □ | ■ | □ | ■ | ■ | ■ | ■ | □ | ■ |
| 24. | □ | ■ | □ | ■ | ■ | ■ | ■ | □ | ■ |
| 25. | □ | ■ | □ | ■ | ■ | ■ | ■ | □ | ■ |
| 26. | □ | ■ | □ | ⊞ | ■ | ■ | ■ | □ | ■ |

Table 1: Comparison of coeval approaches for etablishing roles at runtime based on 26 classifying features extracted from the literature [9,11]. It differentiates between fully (■), partly (⊞) and not supported (□) features.

---

[1] See: `github.com/max-leuthaeuser/SCROLL`
[2] See: `scala-lang.org/api/current/#scala.Dynamic`

```
1   class Person(val name: String)
2   class Car(val licenseID: String)
3   class Location(val name: String)
4
5   class Transportation() extends Compartment {
6     object AutonomousTransport extends Compartment {
7       class SmartCar() {
8         def drive() {
9           info("I am driving autonomously!")
10        }
11      }
12      class Passenger() {
13        def brake() {
14          info(s"I can't reach the brake. I am ${+this name}
                 ↪ and just a passenger!")
15        }
16      }
17    }
18
19    object ManualTransport extends Compartment {
20      class NormalCar() {
21        def drive() {
22          info(s"I am driving with a driver called
                 ↪ ${+one[Driver]() name}.")
23        }
24      }
25      class Driver() {
26        def brake() {
27          info(s"I am ${+this name} and I am hitting the
                 ↪ brakes now!")
28        }
29      }
30    }
31
32    class TransportationRole(source: Source, target: Target) {
33      def travel() {
34        val kindOfTransport = this player match {
35          case ManualTransport => "manual"
36          case AutonomousTransport => "autonomous"
37        }
38        info(s"Doing a $kindOfTransport transportation with
               ↪ the car ${one[Car]().licenseID} from
               ↪ ${+source name} to ${+target name}.")
39      }
40    }
41
42    class Target()
43    class Source()
44  }
```

```
1   val transportation = new Transportation {
2     val peter = new Person("Peter")
3     val harry = new Person("Harry")
4     val googleCar = new Car("A-B-C-001")
5     val toyota = new Car("A-B-C-002")
6
7     new Location("Munich") play new Source()
8     new Location("Berlin") play new Source()
9     new Location("Dresden") play new Target()
10
11    harry play new ManualTransport.Driver()
12    toyota play new ManualTransport.NormalCar()
13
14    +toyota drive()
15    ManualTransport play
16      new TransportationRole(
17        one[Source]("name" ==# "Berlin"),
18        one[Target]()) travel()
19
20    peter play new AutonomousTransport.Passenger()
21    googleCar play new AutonomousTransport.SmartCar()
22
23    +googleCar drive()
24    AutonomousTransport play
25      new TransportationRole(
26        one[Source]("name" ==# "Munich"),
27        one[Target]()) travel()
28
29    +peter brake(); +harry brake()
30  }
```

Fig. 1: The SmartCar example (*instance code*, top) and the corresponding *model code* (left).

```
1   I am driving with a driver called Harry.
2   Doing a manual transportation with the car A-B-C-002 from
        ↪ Berlin to Dresden.
3   I am driving autonomously!
4   Doing a autonomous transportation with the car A-B-C-001
        ↪ from Munich to Dresden.
5   I can't reach the brake. I am Peter and just a passenger!
6   I am Harry and I am hitting the brakes now!
```

Fig. 2: Running the SmartCar example generates the *console output* shown above.

### Acknowledgements

This work is funded by the German Research Foundation (DFG) within the Research Training Group "Role-based Software Infrastructures for continuous-context-sensitive Systems" (GRK 1907).

# A case for Rebel

## A DSL for Product Specifications

Jouke Stoel

CWI, Amsterdam, The Netherlands

jouke.stoel@cwi.nl

## 1. Introduction

Large service organisations like banks have a hard time keeping grips on their software landscape. This is not only visible while performing maintenance on existing applications but also when developing new applications.

One of the problems these organisations face is that they often do not have a clear and uniform descriptions of their products like savings- and current account, loans and mortgages. This makes it hard to reason about changes to existing products and hampers the introduction of new ones. The specifications that *are* written down often contain ambiguities or are out-of-date. Next to this, specification are almost always written down using natural language which is known to lead to numerous deficiencies [1].

To counter these problems we introduce Rebel, a DSL for product specifications. Rebel lets users specify their product in a high-level, unambiguous manner. These specification can then be simulated which enables users to explore their products before they are build.

We have created Rebel for a large Dutch bank and are currently in the process of specifying existing banking products.

Since Rebel is in the early stages of development we would like to use DSLDI to gather feedback on its current design and proposed future directions.

## 2. Rebel

Rebel is a domain specific language for product specifications. It is inspired on formal methods like Z [2], B [3] and Alloy [4]. It is aimed at helping a large Dutch bank in bridging the gap between informal specifications written down in natural language or passed on mouth-to-mouth towards unambiguous, machine interpretable specifications. The main idea behind Rebel is to present the user with a easy to understand syntax and interface while it exploits powerful tooling like verification to check whether the specifications hold under the hood.

Rebel is implemented in RASCAL [5] as a stand-alone DSL.

### 2.1 Requirements

The language needed to fit the following requirements:

- Flexibility - it should be possible to tune it to the problem of the bank we were working with.
- Integration - it should be possible to integrate existing tools like model checkers and connect to existing systems in the banks application landscape.
- Adaptation - it should be easy to learn and the tooling like an IDE should be similar to the tooling currently used.

Considering these requirements we decided to create a new language. This new language needed to be a linguistic hybrid to be able to support both the definition of single products as well as the overlying process.

### 2.2 Design

Rebel is a declarative language and centres around *specifications*. Figure 1 shows an example of such a specification.

A specification describes one product. Specifications contain *fields*, *events*, *invariants* and *life cycle*. Fields declare the data used in the specification. Events describe the possible mutations on the data under certain conditions. Invariants describe global rules which should always hold and life cycle constrains the order of events.

The definition of events and invariants is separated from usage in specifications. This is to promote reuse and to separate the responsibility of implementing an event from using an event in a specification.

Defined fields can only be of built-in types. Events can only reference fields declared in the specification, not fields of other specifications. We made this choice so that the potential state space is smaller when applying verification techniques like model checking.

Events are described using pre- and postconditions. An example event is shown in Figure 2. The semantics are straightforward; if the precondition holds then the postcondition will hold after the event is raised. Events contain runtime instance variables as well as configuration variables. Configuration variables are keyword parameters that can have a default value and can be set when the event is referenced by a particular specification. For instance, the us-

```
specification SavingsAccount {
    fields {
        balance: Time -> Integer
    }

    events {
        openAccount[minimumDeposit=50]
        withdraw[]
        deposit[]
        close[]
    }

    invariants {
        positiveBalance
    }

    lifeCycle {
        initial new -> opened: openAccount
        opened  -> opened: closed: close
        final closed
    }
}
```

**Figure 1.** Example Rebel specification

```
initial event openAccount
    [minimumDeposit : Integer = 0]
    (accountNumber: String, initialDeposit : Integer) {
    preconditions {
        initialDeposit >= minimumDeposit;
    }
    postconditions {
        new this.balance(now) == initialDeposit;
    }
}
```

**Figure 2.** Example of an event definition

age declaration of `openAccount` (Figure 1) sets the event configuration parameter `minimumDeposit` meaning that the `SavingsAccount` uses 50 as a `minimumDeposit` when an account is opened.

Invariants are global rules. They use quantifiers over data to express certain constrains that should always hold. Figure 3 shows an example that states that at all time, saving accounts should have a balance equal to or above zero.

```
invariant positiveBalance {
    all sa:SavingsAccount | all t:Time {
        sa.balance(t) >= 0
    }
}
```

**Figure 3.** Example of an invariant

### 2.3 Simulating specifications

The simulation is aimed at helping product owners and developers gain insight into their specified product. It can be used to check if the specification meets the expectations of the user. Figure 4 shows a screenshot of the simulation of a `SavingsAccount`. The simulation is implemented with the use of the Z3 SMT solver [6].

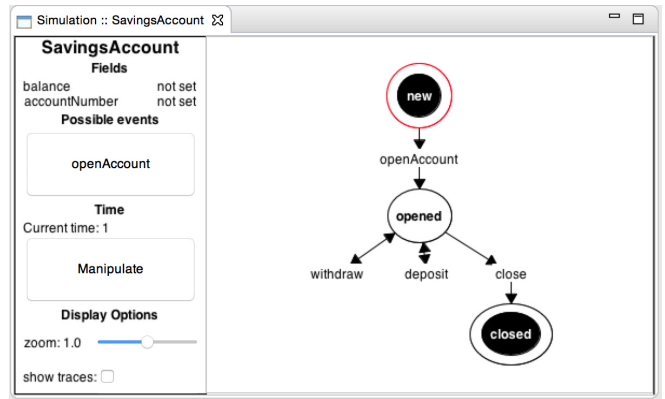

**Figure 4.** Screenshot of simulating the SavingsAccount specification

### 3. Future work

The current version of Rebel supports the definition of single products. Next to this it is also needed to define composition of these products. In other words, the process. Since the specifications only contain fields of built-in types and can only reference themselves it is not possible to compose specifications. To overcome this we propose the use of process algebra [7] for specifying how the individual specification events should be composed. This will give us the ability to specify choices, sequencing, concurrency and communications between specifications. The question will be if we will still be able to reason about (certain parts of) the specifications since composing the specifications will have a large impact on the state space.

An orthogonal aspect is the tooling for Rebel specifications. Next to the simulation we will explore the possibility of model checking. The model checker could be used to find event traces that lead to violations of the invariants. Earlier work has shown that it is possible to translate Rebel specification to Alloy. Alloys analyser was used to find traces which would break the specification. The problem with this approach was scalability. An alternative would be to exploit an SMT solver for the same purpose [8]. One of the challenges here will be how we can bound the data in a smart way to limit the state space.

Ultimately, running systems should be generated from Rebel specifications. Since Rebel is a declarative language it will not always be straightforward to generate a correct system from this. Again SMT solvers might hold the key as shown in other work like [9].

# Flick: A DSL for middleboxes

Network-as-a-Service (NaaS) project[*]

We argue the need for specialised languages to program application-level middleboxes, and describe our design for such a language, through which we seek to make available suitable abstractions for middlebox programming, and constrain what kinds of programs can be expressed.

A *middlebox* is a non-standard router; it carries out a computation on network traffic (beyond decrementing a TTL and recomputing a checksum) and the routing decision on that traffic may be influenced by the computation's result. Examples of middleboxes include firewalls, protocol accelerators, VPN gateways, transcoders, load-balancers and proxies. Middleboxes perform a vital function in today's Internet, in datacentres, and even in home and corporate networks.

An *application-level* middlebox is one that does not deal solely in network-level primitives or protocols, but also (and perhaps exclusively) in application-level protocols and data. Examples include a memcached caching reverse-proxy, and a load-balancer that unsheaths an HTTP/SSL connection before passing the cleartext HTTP data to a backend server.

Middleboxes are usually written in general-purpose languages—usually C. Such languages are sufficiently expressive, enjoy a broad developer population, and a compiler is likely to exist for the intended architecture. It has been observed that even for systems programming, the disadvantages of using a general-purpose language sometimes outweigh the benefits [3,5]. We believe that middlebox programming is an example of this. By their nature, general-purpose languages do not provide suitable abstractions for middlebox programming. Another disadvantage is that general-purpose languages are much too expressive for writing middleboxes, which often do not implement complex behaviour. Indeed middleboxes cannot implement complex behaviour if they are to operate at line rate at high bandwidth.

This suggests the need for a framework in which to write middleboxes. This could be implemented in two ways: as a DSL, or as a library. Either approach could provide more suitable abstractions for middleboxes than is usually provided by a general-purpose language alone. Because of this, either approach could lead to shorter, more readable, code, without significant regression in performance. Work on both approaches has been described in the literature.

DSLs for middlebox programming include Click [4], POF [6], and P4 [2]. All of these are designed for processing packets. (Click can also be regarded as a library, but even there it is oriented towards the implementation of packet processors.) These languages seek to support the use of arbitrary protocols; part of the programmer's task is to encode the packet format.

---

[*]http://www.naas-project.org/

But not all middleboxes are most naturally defined as packet processors—this is particularly the case for *application-level* middleboxes which we seek to support. We can think of middleboxes more generally as processors of arbitrary data extracted from byte streams. xOMB [1] is a library for programming middleboxes in C++. Unlike packet processors, xOMB elements can operate on higher-layer data. But a library does not allow us to impose suitable constraints on programs: using a middlebox library leaves you at liberty to write arbitrary functions in the host language.

We therefore opted to design a DSL for implementing application-level middleboxes. Our language, called Flick, has the following features. First, it is statically typed, and features algebraic types. In addition to being used for type checking and inference, types can be used to synthesise serialisation and deserialisation code for values of a type. This is achieved by decorating a type's declaration with *serialisation annotations*. These specify the precision of integers, for example, and byte ordering for multi-byte values. Second, Flick only allows bounded recursion. It is a Turing-incomplete language; only terminating computations can be expressed in Flick. This is an important constraint that middlebox DSLs enforce, but that libraries cannot, as mentioned earlier. Third, channels and processes are language primitives; such concepts seem native to middleboxes. Channels are typed, and connect processes to other processes, or to external (network) sources or destinations of data. Processes can run concurrently, and contain the middlebox's logic. Fourth, processes encapsulate their own state, but middleboxes may also share state. That is, message-passing is the method used to both notify a process of new data, and to provide that data; but shared memory can be used when notification is not necessary. We found this useful for describing shared caches, such as in the encoding of a load-balancer for the memcached key-value database, shown below.

```
proc Memcached: (cmd/cmd client, [cmd/cmd] backends)
  global cache := empty_dict
  client => test_cache(client, backends, cache)
  backends => update_cache(cache) => client
```

Here the process `Memcached` has a channel `client`, and an array of channels `backends` with which it can communicate with other processes. All channels in this process can yield and accept values of type `cmd`, values of which are memcached commands. `update_cache` and `test_cache` are functions. The body of the process simply forwards requests from `client` to a backend unless the reply has been cached; and it forwards replies from backends to the client, after caching them locally.

So far we have a formal semantics for the core expression language, and a partial compiler to a runtime of our devising. Future work involves language features such as exception handling and resource estimation, as well as extending the compiler to include more targets, including reconfigurable hardware.

# Towards a Next-Generation Parallel Particle-Mesh Language[*]

Sven Karol[1], Pietro Incardona[2,3], Yaser Afshar[2,3], Ivo F. Sbalzarini[2,3], and Jeronimo Castrillon[1]

1   Chair for Compiler Construction, Center for Advancing Electronics Dresden, TU Dresden, Dresden, Germany
    `[sven.karol|jeronimo.castrillon]@tu-dresden.de`
2   MOSAIC Group, Chair of Scientific Computing for Systems Biology, Faculty of Computer Science, TU Dresden, Dresden, Germany
3   Center for Systems Biology Dresden, Max Planck Institute of Molecular Cell Biology and Genetics, Dresden, Germany
    `[afshar|incardon|ivos]@mpi-cbg.de`

## Abstract

We present our previous and current work on the parallel particle-mesh language PPML—a DSL for parallel numerical simulations using particle methods and hybrid particle-mesh methods in scientific computing.

## 1    Introduction

During the past years, domain-specific languages (DSLs) gained central importance in scientific high-performance computing (HPC). This is due to the trend towards HPC clusters with heterogeneous hardware—today, mainly using multi-core CPUs as well as streaming processors such as GPUs—in the future, using many-core CPUs, and potentially also reconfigurable processors or data-flow processing units. Writing programs for these machines is a challenging and time-consuming task for scientific programers, who do not only need to develop efficient parallel algorithms for the specific problem at hand, but also need to tune their implementations in order to take advantage of the cluster's hardware performance. This does not only require experience in parallel programming, e.g. using OpenMP, OpenCL, or MPI, but also in computer architectures and numerical simulation methods, leading to the so-called "knowledge gap" in program efficiency [12]. Besides, it renders the simulation codes hardly portable. DSLs can help two-fold: First, they allow scientific programmers to write programs using abstractions closer to the original mathematical representation, e.g., partial differential equations. Second, they transparently encapsulate hardware-specific knowledge.

In the proposed talk, we focus on the parallel particle-mesh language (PPML) [3]. This language provides a macro-based frontend to the underlying PPM library [13, 2] as a parallel run-time system. We analyze PPML's implementation as well as its advantages and disadvantages w.r.t. state-of-the-art DSL implementation techniques. Based on this analysis, we discuss our early efforts in realizing the next version of PPML (Next-PPML) in conjunction with a redesign of the PPM library in C++.

[*] This work is partially supported by the German Research Foundation (DFG) within the Cluster of Excellence "Center for Advancing Electronics Dresden".

## 2    Particle and Mesh Abstractions

In scientific computing, discrete models are naturally simulated using particles that directly represent the discrete entities of the model, such as atoms in a molecular-dynamics simulation or cars in a traffic simulation. These particles carry properties and interact with each other in order to determine the evolution of these properties and of their spatial location. But also continuous models, written as partial differential equations, can be simulated using particles. In this case, the particle interactions discretize differential operators, such as in the DC-PSE method [14]. This is often combined and complemented with mesh-based discretizations, such as the finite-difference method. Mesh and particle discretizations are equivalent in that they approximate the simulated system by a finite number of discrete degrees of freedom that are the particles or the mesh cells. When using particles together with meshes, it is sufficient to consider regular Cartesian (i.e., checkerboard) meshes, since all irregular and sub-grid phenomena are represented on the particles, which can arbitrarily distribute in the domain.

Particles and meshes hence define data abstractions. A particle is a point abstraction that associates a location in space with arbitrary properties, like color, age, or the value of a continuous field at that location. These properties, as well as the particle locations, are updated at discrete time steps over the simulation period by computing interactions with surrounding particles within a given cut-off radius. Meshes are topological abstractions with defined neighborhood relations between cells. The properties are stored either on the mesh nodes or the mesh cells. The PPM library supports both types of abstractions, and also provides conversion operators between them (i.e., particle-mesh interpolation).

## 3    Current Status of PPM(L)

Currently, PPM is implemented in object-oriented Fortran2003 [13, 2] and PPML is a macro system embedded in Fortran2003 [3]. PPML and PPM support transparently distributed mesh and particle abstractions, as well as parallel operations over them. This also includes properties and iterators. Different domain-decomposition algorithms allow for the automatic distribution of data over the nodes of an HPC cluster. Assigning data and work to processing elements is automatically done by a graph partitioning algorithm, and communication between processing elements in transparently handled by PPM "mappings". The mathematical equations of the model to be simulated are written in LaTeX-like math notation with additional support for differential operators and dedicated integration loops [3].

Syntactically, PPML is an extension of Fortran2003 providing the aforementioned abstractions as domain-specific language concepts. Technically, the language is implemented as a source-to-source transformation relying on a mixture of macro preprocessing steps where macro calls are interspersed with standard Fortran2003 code. Besides C-style preprocessor directives, PPML also supports non-local macros. These are implemented in Ruby using eRuby as a macro language and the ANTLR parser generator for recognizing macro output locations, such as integration loops. Hence, PPML is partially realized using an island grammar [11].

In this preliminary form, PPML has already nicely demonstrated the benefits of embedded DSLs for scientific HPC. It has reduced both the size and the development time of scientific simulation codes by orders of magnitude [3]. It hides much of the parallelization intricacies (PPML automatically generates MPI) from scientific programmers without preventing them from using all features of the underlying programming language. The latter is essential since a DSL may not cover all potential corner cases, and may not always deliver top performance. However, the current light-weight implementation of PPML has severe disadvantages when it comes to code analysis algorithms targeting the whole program and domain-specific optimizations based thereon. Moreover, PPML programs are difficult to debug due to a lack of semantic error messages. We hence present our intended improvements addressing these issues in Next-PPML.

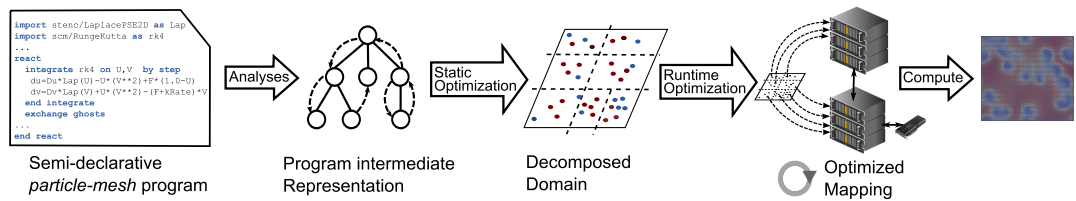

**Figure 1** Compiler and grammarware-based language processing chain of Next-PPML.

## 4 Approach to Next-PPML

Next-PPML is a language extension using grammarware and compilerware. This allows us to analyze larger portions of the program code. Examples such as the universal form language (UFL) [1] for finite-element meshes, the Liszt language for mesh-based solvers [7], and the Blitz++ [15] stencil template library have shown that domain-specific analyses and built-in abstractions are beneficial for scientific computing DSLs. Hence, similar concepts will be considered in the Next-PPML language.

Figure 1 conceptually illustrates the planned tool chain. First, the embedded DSL program is parsed to an AST-based intermediate representation. This representation already contains control-flow edges. After computing domain-specific static optimizations on this intermediate representation, including optimizations to the communication pattern of the parallel program, the Next-PPML compiler generates an executable (or source code) which is then used to run the simulation on a parallel HPC cluster. During the simulation run, the application continuously self-optimizes, e.g., for dynamic load balancing. While static optimizations are handled by the DSL compiler, dynamic runtime optimization are handled by the PPM library, which may rely on information provided by the DSL program.

Ideally, the new language uses a declarative approach that bases on an existing programming-language grammar and extends it with new productions. Some well-known candidates for this are Stratego/XT [4], TXL [6], JastAdd [8] or EMFText [9]. However, the target language is C++11 which has no simple declarative specification. Hence, it is difficult to estimate if the above-mentioned tools would scale, and implementing a C++ frontend is a huge project on its own. Therefore, we prefer Clang [10] as an implementation framework, which already provides built-in analyses that can be adopted and extended.

## 5 Conclusions

Hybrid particle-mesh simulations are the only scientific computing framework that is able to simulate models of all four kingdoms: continuous/deterministic, continuous/stochastic, discrete/deterministic, and discrete/stochastic. This versatility makes the hybrid particle-mesh paradigm a prime target for a generic parallel HPC DSL for scientific computing. Prior work has shown the power of parallelization middleware libraries like PPM, and embedded DSLs like PPML. Over the past 10 years, they have reduced code development times for parallel scientific simulations from years to hours, and enabled unprecedented scalability on large HPC machines [5]. The envisioned Next-PPML will address current shortcomings in code generation, static and dynamic optimization, and semantic error checking and reporting. It is co-developed with the next generation of the PPM library in C++ using a semi-declarative language design.

# DSLs for Graph Algorithms and Graph Pattern Matching

Proposal for DSLDI 2015 - submitted 8 April 2015

*Oracle Labs*

Oskar van Rest
Sungpack Hong
Hassan Chafi

While SQL and procedural languages like PL/SQL have proven very successful for relational-data-based applications, the increasing importance of graph-data-based applications is fueling the need for graph processing DSLs. On the one hand is the need for a procedural-like language that allows you to implement graph algorithms, like Dijkstra's shortest path algorithm and Brandes' Betweenness Centrality algorithm, while on the other hand is the need for a declarative graph pattern-matching DSL, much like SQL for graphs. We believe that our DSLs Green-Marl and GMQL are tailored for these two types of graph workloads and we aim at making the languages standards in the field of graph processing.

Green-Marl is a procedural DSL for implementing graph algorithms for use cases such as product recommendation, influencer identification and community detection. The language allows you to intuitively define a wide set of graph algorithms by providing 1) graph-specific primitives like nodes and edges, 2) built-in graph traversals like BFS and DFS, and 3) built-in iterations over neighbors and incoming neighbors of nodes in the graph. Furthermore, Green-Marl algorithms can be automatically translated into parallel implementations in C++, Java or other general purpose languages. Thus, Green-Marl users can intuitively define their graph algorithms using high-level graph constructs, and then run their algorithms efficiently on large graphs.

Our other language, GMQL, is a declarative graph pattern-matching query language that borrows syntax from Neo4j's Cypher and from the SPARQL query language. It is like SQL for graphs, but instead allows you to query using concepts like nodes, edges and paths, instead of tables, rows and columns. A graph pattern can be defined in the form of a set of nodes that are connected via edges or arbitrary-length paths together with a set of constraints on (properties) of the nodes, edges and paths. More complicated patterns that are typical for query languages, such as negation and optional matching, are also supported. The execution of a GMQL query comes down to finding all the instances of the specified pattern inside the graph. Again, execution can be done in parallel to support the efficient processing of large graphs.

The Green-Marl and GMQL compilers that we created, target our efficient, parallel, and in-memory graph analytic framework PGX. PGX supports loading graphs from various popular flat file graph formats and can also load graphs from an Oracle database. Graphs are efficiently stored in memory using the Compressed Sparse Row (CSR) format. This format allows for huge graphs to fit into the memory of a single machine, in such a way that they can be processed efficiently. Our framework is also very portable: The Green-Marl and GMQL compilers target both our Java-based and C++-based PGX backbends. Furthermore, we are working on a distributed backend that allows for the processing of even bigger graphs that do not fit into the memory of a single machine. In such distributed mode, graphs are partitioned across multiple machines and data is exchanged between machines using high-speed Infiniband or 10G-E.

So far, we have seen great performance improvements with PGX, when comparing it to other graph frameworks: Green-Marl algorithms are processed one to two orders of magnitude faster than corresponding implementations with the popular machine learning framework GraphLab. Furthermore, GMQL queries are processed two to four orders of magnitude faster than corresponding Cypher queries with Neo4j.

Because our PGX framework can process large graphs very efficiently, we decided to also create support for the popular RDF graph data model. This means we import RDF graphs into PGX and query them using either GMQL or the standard RDF query language SPARQL. In order to import RDF graphs, we created a translation from RDF graphs into PGX' property graphs. Furthermore, in order to query using SPARQL, we have created a translation from SPARQL to GMQL. This translation currently supports a subset of W3C's SPARQL 1.1.

The Green-Marl, GMQL and SPARQL compilers were implemented using the Spoofax language workbench. Spoofax provides high-level DSLs for specifying grammars, name binding, type systems and transformations. By using these DSLs, we can keep our code base small and maintainable, while we obtain much compiler functionality, such Eclipse IDE integration, from Spoofax for free.

Our Spoofax-based SPARQL implementation initially functioned merely as a building block for the SPARQL-to-GMQL translation. However, the Eclipse editor that we obtained from Spoofax has become a product on its own and is useful for anyone who wants to write SPARQL queries. The editor has full support for W3C's SPARQL 1.1 and provides editor features like formatting, code completion, syntax checking, name-based checks for variables and prefixes, and editor navigation from name uses to their definitions. Many of the features came with little implementation effort. For example, our SPARQL grammar definition in Spoofax' grammar definition formalism SDF3, gave us a parser, a formatter, syntax-checks and error recovery rules, while our SPARQL name-binding definition in Spoofax' name binding language NaBL gave us name-based code completion, name checks and editor navigation. We were even able to encode the more complicated name-binding rules needed for SPARQL constructs like NOT EXISTS, MINUS and for Subqueries in NaBL, in very few lines of code.

Finally, we also allow our Green-Marl, GMQL and SPARQL compilers to be linked against a particular graph such that we can perform additional compile-time error checking. For example, our GMQL compiler is able to warn a user when they have misspelled a property name, based on the node and edge properties that are available in a particular graph. We also provide such kinds of checks when querying schema-less RDF graphs, by extracting schema information from an RDF graphs when it is loaded into PGX. In our Eclipse editors, such error checking even happens in real-time.

# DSLs of Mathematics, Theorems and Translations

Cezar Ionescu*   Patrik Jansson*

In this talk, we present some of the ideas behind the course on *DSLs of Mathematics (DSLM)*, currently in preparation in Chalmers.

We view mathematics as a rich source of examples of DSLs. For example, the language of group theory, or the language of probability theory, embedded in that of measure theory. The idea that the various branches of mathematics are in fact DSLs embedded in the "general purpose language" of set theory was (even if not expressed in these words) the driving idea of the Bourbaki project, which exerted an enormous influence on present day mathematics.

In DSLM, we consequently develop this point of view, aiming to show computer science students that they can use the tools from software engineering and functional programming in order to deal with the classical continuous mathematics they encounter later in their studies.

In this talk, we'll start with the simple example of the standard development of a calculus of derivatives. This can be seen as a DSL whose semantics are given in terms of limits of real sequences. We can try to give alternative semantics to this language, in terms of complex numbers. This leads to the notion of holomorphic function, and to an essentially different calculus than in the real case.

Our second example is that of extending the language of polynomials to power series. This DSL can also be interpreted in various domains: real numbers, complex numbers, or intervals.

In the case of complex numbers, a fundamental theorem creates a bridge between the DSL of derivatives and that of power series, through the identity of holomorphic and (regular) analytic functions. This leads to the discussion of translation between DSLs, an aspect which is fundamental in mathematics, but has been somewhat neglected by computer science. Thus, we believe that a closer examination of the DSLs of mathematics can also be relevant for practical software engineering.

*Chalmers University of Technology

# Check Syntax: An Out-of-the-Box Tool for Macro-Based DSLs

Spencer Florence  Ryan Culpepper  Matthew Flatt  Robert Bruce Findler
Northwestern University  Northeastern University  University of Utah  Northwestern University

## Abstract
Racket supports DSL construction through an open compiler, where the compiler is made open through its macro system. Programmers define new syntactic constructs by macro-expansion to existing constructs, using the Racket module system to hide the originals and rename the new ones as needed. Taken together, these facilities permit the definition of new languages, even enabling new languages to give new semantics to familiar syntax.

In this world, all programs are compiled to a well-known intermediate language, and tools can operate on that structure to compute information about programs—including programs written in a DSL that the tool knows nothing about. Crucially, binding information and original source locations are available in the intermediate language. This information allows tools to provide keybindings to hop between bound and binding occurrences of identifiers and rename them. Since Racket's documentation system is based on binding, tools can also conveniently access API documentation.

## 1. Building a Macro-based Language in Racket

Programming tools typically support multiple languages by compiling all programs into a shared intermediate language. In Racket, the representation of the intermediate language is the same as for macro transformations: an enriched form of S-expressions that embeds source-location and binding information. This representation offers an especially convenient way to make new languages via macros, where the new languages inherit tools that operate on macro-expanded programs.

For example, to build a DSL that uses call-by-need evaluation instead of Racket's usual call-by-value convention, we start with a `wait` construct for delaying the evaluation of an expression. The `wait` form is defined using `define-syntax-rule`, which directs the compiler to replace expressions matching one subexpression with another.

```
(define-syntax-rule
  (wait e)
  (Wait #t (λ () e)))
```

The definition of `wait` specifies that the body `e` in any `(wait e)` is wrapped in a thunk (thus delaying the evaluation of `e`) and the thunk is packaged into a `Wait` structure:

```
(struct Wait (waiting? TorV) #:mutable)
```

This `struct` declaration indicates that the `Wait` structure has two mutable fields, `waiting?` and `TorV`, and it defines supporting functions: `Wait-waiting?`, which accepts a `Wait` structure and returns the value in the first field; `set-Wait-waiting?!`, which accepts a `Wait` structure and a value and a new `waiting?` field value and mutates the structure; `Wait-TorV`, which returns the value of a `Wait` structure's second field; and `set-Wait-TorV!`, which mutates the value of a `Wait` structure's second field.

To force a delayed evaluation, we define an `act` function. It operates on a `Wait` structure, returning the result of the thunk encapsulated in the second field and caching the thunk's result.

```
(define (act w)
  (cond
    [(Wait? w)
     (when (Wait-waiting? w)
       (set-Wait-TorV! w ((Wait-TorV w)))
       (set-Wait-waiting?! w #f))
     (act (Wait-TorV w))]
    [else w]))
```

Next, we define macros that compile the various constructs of our new language into constructs that already exist in Racket. In this case, we'll use the `wait` functionality and exploit Racket's λ, a conditional form, and some simple arithmetic operations, lifting them into our lazy language. The following 8 lines provide the rest of a lazy language implementation, using a mixture of function definitions and macros.

```
(define-syntax-rule
  (app f x ...)
  ((act f) (wait x) ...))
(define (multiply a b) (* (act a) (act b)))
(define (subtract a b) (- (act a) (act b)))
(define-syntax-rule
  (if0 e1 e2 e3)
  (if (= (act e1) 0) e2 e3))
```

Using these constructs directly would be awkward. They do not have the right names, and it is easy to accidentally step outside of the language and use, for example, Racket's `*` operation instead of the `multiply`. To avoid those problems, we put the definitions into a module, we use `provide`'s rename-on-export capability to provide those operations with their expected names, and we hide everything else (by simply not providing anything else). The only subtle point here is that, when the Racket macro expander sees an open parenthesis with no macro following it, it inserts a reference to the `#%app` macro to make function application explicit, so we must export `app` as `#%app` to cooperate with this feature of macro expansion. Then, we can install the language as a Racket package, so that if the first line of a file is `#lang mini-lazy`, the rest of the file sees our bindings. For example, running the following program (using the call-by-name Y combinator) produces 3628800.

```
#lang mini-lazy
(((λ (f) ((λ (x)
            (f (x x)))
          (λ (x)
            (f (x x)))))
  (λ (fac)
    (λ (n)
      (if0 n 1 (* n (fac (- n 1))))))
  10)
```

Even though this language has a radically different order of evaluation from Racket, it still compiles into Racket and, even better, scope is preserved by this compilation.

## 2. Check Syntax

Check Syntax runs continuously as part of the DrRacket IDE. Each time the user edits a program, Check Syntax takes the content as a string, parses it using the current language's parser, and hands it off to the macro system, resulting in a program where the macros have all been expanded away. In this state, programs are in a well-known language that contains functions, conditionals, variable binding, variable reference, and other core forms of Racket.[1]

Check Syntax traverses a program's expansion, searching for bindings and references. The expanded form of a program is an enriched form of S-expressions known as *syntax objects*. A syntax object includes information about whether or not a form appeared in the original program and its location in the source, and identifiers extracted from an expanded program can be compared to determine whether they refer to the same binding. Check Syntax collects this information into pairs of source locations, which DrRacket uses to draw binding information in the editor window.

For the example at the end of the previous section, these are the set of arrows that Check Syntax collects for the lexical variables:[2]

```
#lang mini-lazy
(((λ (f) ((λ (x)
            (f (x x)))
          (λ (x)
            (f (x x)))))
  (λ (fac)
    (λ (n)
      (if0 n 1 (* n (fac (- n 1))))))
  10)
```

Check Syntax uses this information in several ways. First, when the mouse is over a variable occurrence, Check Syntax draws arrows to the other occurrences. Second, Check Syntax provides key-bindings to jump between the occurrences of a variable in a file. Finally, Check Syntax offers a bound-variable rename facility. The user selects a variable and supplies a new name, and Check Syntax follows the arrows to rename the variables.

The technique of traversing expanded programs can support a myriad of different languages, even parenthesis-challenged languages like Datalog:

```
#lang datalog
ancestor(A, B) :-
    parent(A, B).
ancestor(A, B) :-
    parent(A, C), ancestor(C, B).
```

---

[1] `http://docs.racket-lang.org/reference/syntax-model.html#%28part._fully-expanded%29`

[2] Naturally, the figures in this paper are generated by running Check Syntax on the shown program text as the paper is typeset.

and Algol 60:

```
#lang algol60
begin
  integer procedure SIGMA(x, i, n);
    value n; integer x, i, n;
  begin
    integer sum;
    sum:=0;
    for i:=1 step 1 until n do
      sum:=sum+x;
    SIGMA:=sum;
  end;
  integer q;
  println(SIGMA(q*2-1, q, 7));
end
```

## 3. Scaling Up

For many constructs, the process described in the previous section is enough to reconstruct the binding structure of a program. Other constructs appear to the programmer as binding forms and variable references, but they are compiled into data-structure accesses. For example, one of the modularity constructs in Racket, unit, turns variable definitions into the creation of a pointer and variable references into pointer dereferences. Similarly, the pattern-matching part of `define-syntax-rule` turns variable references into function calls that destructure syntax objects.[3]

To handle such forms, Check Syntax needs a little cooperation from the implementing macro. When producing an expanded form, a macro can add properties to the result syntax objects. These properties have no effect on how the code runs, but they are used by Check Syntax to draw additional arrows. More precisely, the macro can add a property indicating that some syntax object (i.e., a piece of the input to the transformation) was a conceptually a binding or a reference; Check Syntax uses those syntax objects for renaming and navigation, just like the program's other syntax objects.

Check Syntax also recognizes properties for tool-tip information. A macro can put an arbitrary string as a property on the result of a macro, and Check Syntax displays the string in a tooltip. For example, Typed Racket adds properties so that Check Syntax displays the types of expressions.

In Racket, the build process for documentation creates a database that maps module names and exports to the file containing the documentation and to a function's contract or a syntactic form's grammar. Check Syntax can examine a variable in the fully expanded program and discover which module exported the variable. It uses that information to consult the database and can render the contract/grammar directly in the DrRacket window, including a link to the full documentation.

In the following Typed Racket program, highlighting shows the locations where Check Syntax finds documentation in the database:

```
#lang typed/racket
(: fib (-> Integer Integer))
(define (fib n)
  (cond
    [(= n 0) 0]
    [(= n 1) 1]
    [else (+ (fib (- n 1))
             (fib (- n 2)))]))
```

**Acknowledgments.** Thanks to Matthias Felleisen for comments on earlier drafts and feedback on Check Syntax over the years.

---

[3] The `define-syntax-rule` form is itself is a macro that expands into the primitive macro-building form with a compile-time syntax transformer that is synthesized from the specified pattern and template.

# Dynamic Compilation of DSLs

Vojin Jovanovic and Martin Odersky

EPFL, Switzerland
{firstname}.{lastname}@epfl.ch

Domain-specific language (DSL) compilers use *domain knowledge* to perform *domain-specific optimizations* that can yield several orders of magnitude speedups [4]. These optimizations, however, often require knowledge of values known only at program runtime. For example, in matrix-chain multiplication, knowing matrix sizes allows choosing the optimal multiplication order [2, Ch. 15.2] and in relational algebra knowing relation sizes is necessary for choosing the right join order [6]. Consider the example of matrix-chain multiplication:

```
val (m1, m2, m3) = ... // matrices of unknown size
m1 * m2 * m3
```

In this program, without knowing the matrix sizes, the DSL compiler can not determine the optimal order of multiplications. There are two possible orders `(m1*m2)*m3` with an estimated cost `c1` and `m1*(m2*m3)` with an estimated cost `c2` where:

```
c1 = m1.rows*m1.columns*m2.columns+m1.rows*m2.columns*m3.rows
c2 = m2.rows*m2.columns*m3.columns+m1.rows*m2.rows*m3.columns
```

Ideally we would change the multiplication order at runtime only when the condition `c1 > c2` changes. For this task *dynamic compilation* [1] seems ideal.

Yet, dynamic compilation systems—such as DyC [3] and JIT compilers—have shortcomings. They use runtime information primarily for specialization. In these systems profiling tracks *stability* of values in the user program. Then, *recompilation guards* and *code caches* are based on checking equality of current values and previously stable values.

To perform domain-specific optimizations we must check stability, introduce guards, and code caches, based on the computation specified in the DSL optimizer—outside the user program. Ideally, the DSL optimizer should be agnostic of the fact that input values are collected at runtime. In the example stability is only required for the condition `c1 > c2`, while the values `c1` and `c2` themselves are allowed to be unstable. Finally, recompilation guards and code caches would recompile and reclaim code based on the same condition.

An exception to existing dynamic compilation systems are Truffle [7] and Lancet [5]. They allow creation of user defined recompilation guards. However, with Truffle, language designers do not have the full view of the program, and thus, can not perform global optimizations (e.g., matrix-chain multiplication optimization). Further, recompilation guards must be manually introduced and the code in the DSL optimizer must be modified to specially handle decisions based on runtime values.

We propose a dynamic compilation system aimed for domain specific languages where:

- DSL authors declaratively, at the definition site, state the values that are of interest for dynamic compilation (e.g., array and matrix sizes, vector and matrix sparsity). These values can regularly be used for making compilation decisions throughout the DSL compilation pipeline.
- The instrumented DSL compiler transparently reifies all computations on the runtime values that will affect compilation decisions. In our example, the compiler reifies and stores all computations on runtime values in the unmodified dynamic programming algorithm [2] for determining the optimal multiplication order (i.e., `c1 > c2`).
- Recompilation guards are introduced automatically based on the stored DSL compilation process. In the example the recompilation guard would be `c1 > c2`.
- Code caches are automatically managed and addressed with outcomes of the DSL compilation decisions instead of stable values from user programs. In the example the code cache would have two entries addressed with a single boolean value computed with `c1 > c2`.

The goal of this talk is to foster discussion on the new approach to dynamic compilation with focus on different policies for automatic introduction of recompilation guards: *i)* heuristic, *ii)* DSL author specified, and *iii)* based on domain knowledge.

# A practical theory of language-integrated query
# —and—
# Everything old is new again

Philip Wadler
The University of Edinburgh
`wadler@inf.ed.ac.uk`

15 April 2015

The proposed talk would consist of summaries of two recent pieces of work, both concerned with applying quotation and the subformula property to domain-specific languages. Both will also be the subject of an invited talk by the author at Curry On.

**A practical theory of language-integrated query**
James Cheney, Sam Lindley, Philip Wadler
submitted to ICFP 2015
`http://homepages.inf.ed.ac.uk/wadler/topics/recent.html#essence-of-linq`

> Language-integrated query is receiving renewed attention, in part because of its support through Microsoft's LINQ framework. We present a theory of language-integrated query based on quotation and normalisation of quoted terms. Our technique supports abstraction over values and predicates, composition of queries, dynamic generation of queries, and queries with nested intermediate data. Higher-order features prove useful even for constructing first-order queries. We prove that normalisation always succeeds in translating any query of flat relation type to SQL. We present experimental results confirming our technique works, even in situations where Microsoft's LINQ framework either fails to produce an SQL query or, in one case, produces an avalanche of SQL queries.

**Everything old is new again: Quoted Domain Specific Languages.**
Shayan Najd, Sam Lindley, Josef Svenningsson, Philip Wadler
ICFP 2013
`http://homepages.inf.ed.ac.uk/wadler/topics/recent.html#qdsl`

> We describe a new approach to domain specific languages (DSLs), called Quoted DSLs (QDSLs), that resurrects two old ideas: quotation, from McCarthy's Lisp of 1960, and the subformula property, from Gentzen's natural deduction of 1935. Quoted terms allow the DSL to share the syntax and type system of the host language. Normalising quoted terms ensures the subformula property, which guarantees that one can use higher-order types in the source while guaranteeing first-order types in the target, and enables using types to guide fusion. We test our ideas by re-implementing Feldspar, which was originally implemented as an Embedded DSL (EDSL), as a QDSL; and we compare the QDSL and EDSL variants.

# Panel Discussion

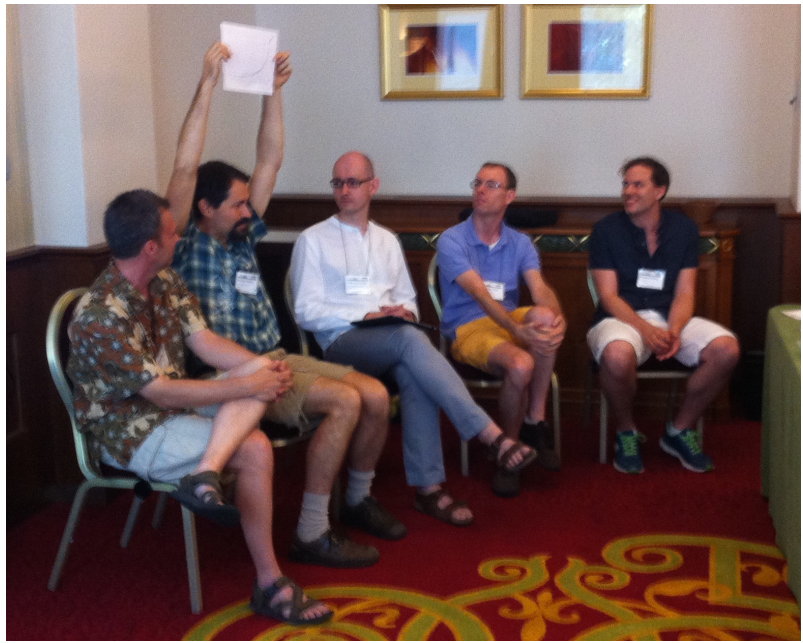

Figure 1: Discussion Panel at DSLDI'15. From left to right: Matthew Flatt, Jonathan Aldrich, Andrzej Wąsowski Laurence Tratt and Sebastian Erdweg

### Position Statements of the Panelists

**Jonathan Aldrich**  Position: DSL frameworks should guarantee the absence of syntactic conflict, and support unanticipated interoperation between DSLs in the code and during typechecking, execution, and debugging, without losing aspects that are special to each DSL.

Rationale: Programming today is all about composition; developers gain enormous leverage from libraries, and expect them to work together even if they were designed separately. Conflicts that prevent compilation when you merely import two different DSLs are completely unacceptable in this world. Furthermore, most of the value from DSLs comes when they work like "real" languages, with checking, execution, and debugging facilities that are natural; an 80% solution is not going to convince most real-world developers to adopt a DSL. In a composition-based world, therefore, all these facilities must work even when multiple DSLs are used together.

Concrete Illustration: Here's a multi-part challenge problem for language composition in DSL frameworks:

- (A) Have different developers independently design and build DSLs for state machines and structured synchronous programming

- (B) The DSL framework should guarantee that these DSLs can be used together without having to resolve any syntactic conflicts

- (C) Write a state machine and a structured synchronous program that drives its state transitions, ideally with no visible role played by the DSL framework.

- (D) Statically verify that the structured synchronous program does not misuse the state machine (e.g. by generating transitions that aren't appropriate for the machine's state)

- (E) With respect to task D, report any errors in a way that is consistent with both the structured synchronous program and the state machine (rather than some translation of each).

The two developers in A are not allowed to communicate or to anticipate tasks B-E. Tasks B–E must be done without changing the DSLs developed in A.

**Sebastian Erdweg**  The same language features reoccur in the design of many DSLs: operations on primitive data, operations on structured data, conditionals and backtracking, error handling, and many more. Yet, we have no principled way of composing basic language blocks into working DSLs and we have no way of detecting and eliminating interactions between language features. This is one of the big open challenges in the area of DSLs.

**Matthew Flatt**  How can language-composition tools mediate extensions that depend on (or, alternatively, adapt to) different semantics of shared constructs, such as function application?

For example, what happens when a form whose implementation depends on eager evaluation is used in an otherwise lazy context? Or what happens when a from that implies a function application is used in a language where function application is meant to be syntactically restricted to first-order functions?

Racket's hygienic-macro approach reflects core constructs like function application through macros, such as `#%app`, whose use is typically implicit. A macro by default adopts its definition-site implementation of such macros, which is usually the right approach. That means, however, that a macro that uses eager function application has questionable behavior in a lazy use context. Similarly, macros tend not to respect the function-application constraints of a context like Beginner Student Language. A macro can adapt to a use-site notion of `#%app` by non-hygienically referencing `#%app` from the use context, but that approach is relatively tedious not not commonly followed.

**Laurence Tratt**  Language composition challenge: *Integrating existing languages into a language composition framework.* Controversial statement: *People have not shown themselves hugely interested in the forms of language composition we've given them thus far.*

**Andrzej Wąsowski**

- I believe that language composition is not a language problem, but a software engineering problem.

- If you are lucky then the DSLs are composed by framework designers, which are usually means very good programmers (or at least above-average).

- More often languages are composed by framework *users*, who design systems (often average programmers or worse). You rarely find serious project using less than 5 languages, and 20 is a norm. Many of them DSLs.

- The challenge is how to allow language composition (or integration) for non-language designers, but for system designers (the language users), so that they still get static checking, meaningful messages, across language testing, etc.

**Summary of the Discussion**

- The practical value of a DSL does not grow linearly with the quality of the implementation. You need a to have really polished framework/ecosystem to deliver value. (80/20 or 20/80 (todo)).

- The software engineering perspective on language composition is different from the programming language perspective. The software engineering perspective emphasizes language interopability, cross-language IDE support, and DSLs that are not necessarily like programming languages (e.g., config files, build files, deployment descriptors, data mapping files, etc.).

- Is using different languages in the same file an essential aspect of language composition? Is language interopability an instance of language composition, and what are the consequences for performance?

- Translating all language features down to a single virtual machine for execution, is possible, but there can be costs in terms of performance. Example: the JVM is a state-of-the-art virtual machine, but is not suitable for executing Prolog.

- Modular language components seems to be an extremely hard to achieve goal, but it is necessary at the same time. Without it, the number of feature interactions quickly explodes.

- There is a need for language interfaces: what features, services or constructs are exported from a language component?

- Integrating syntax and semantics are only two language aspects that need to be composed for a realistic programming experience of a composed language. Examples include: name binding, type checking, IDE features, etc.

- Cross language name analysis seems feasible and would solve real problems of programmers, now. A successful example is Jetbrains' IntelliJ which integrates all references in Web framework configuration files with Java.